# Surface plasmon resonance study of the actin-myosin sarcomeric complex and tubulin dimers

Hans A. Schuessler[1], Alexander A. Kolomenskii[1], Andreas Mershin [1,*], and D. V. Nanopoulos [1,2,3]

[1] Department of Physics, Texas A&M University, College Station, TX 7843-4242, USA

[2] Astro Particle Physics Group, Houston Advanced Research Center (HARC), The Mitchell Campus, Woodlands, TX 77381, USA

[3] Academy of Athens, Chair of Theoretical Physics, Division of Natural Sciences, 28 Panepistimiou Avenue, Athens 10679, Greece

**Abstract**

Biosensors based on the principle of surface plasmon resonance (SPR) detection were used to measure biomolecular interactions in sarcomeres and changes of the dielectric constant of tubulin samples with varying concentration. At SPR, photons of laser light efficiently excite surface plasmons propagating along a metal (gold) film. This resonance manifests itself as a sharp minimum in the reflection of the incident laser light and occurs at a characteristic angle. The dependence of the SPR angle on the dielectric permittivity of the sample medium adjacent to the gold film allows the monitoring of molecular interactions at the surface. We present results of measurements of cross-bridge attachment/detachment within intact mouse heart muscle sarcomeres and measurements on bovine tubulin molecules pertinent to cytoskeletal signal transduction models.

[*] Corresponding author: mershin@physics.tamu.edu

## 1. Introduction

The technique of surface plasmon resonance (SPR) [1,2] allows measurement of changes in the optical properties of a medium adjacent to a thin metal film. Practical applications of the SPR method include chemical sensors [3,4] and biosensors [5]. Specifically, the SPR technique is by now a well-established method for the analysis of interactions among biomolecules [6]. SPR curves can be measured either by varying the angle or the wavelength of the incident light [7-9]. In this paper we discuss our application of the SPR technique to monitoring activity of mouse cardiac muscle sarcomeres [10] and measurement of the dielectric properties of the bovine cytoskeletal protein tubulin.

A surface plasmon (SP) is an electromagnetic wave that can propagate along the surface of a dielectric-metal interface [2]. Surface plasmons can be excited by shining light on a layered system consisting of a transparent medium on one side, a metal film (most often gold or silver) and a dielectric on the other. When the light is incident at an angle greater than the critical angle of total internal reflection, an evanescent wave is produced and penetrates into the adjacent medium. The maximum coupling between the evanescent wave and the surface plasmon takes place when their phase velocities coincide at which point the surface plasmon is excited at resonance. Thus, the surface plasmon resonance (SPR) occurs at a characteristic angle of incidence. This angle depends on the thickness as well as the dielectric permittivities of the layers of the adjacent media. Since the permittivities depend on the frequency of the exciting laser light, the resonance angle does too. The most convenient geometry for the development of a sensor is the Kretschmann-Raether configuration which consists of a glass prism, a metal film and the adjacent medium that is to be probed (figure 1).

## 2. Biomolecular polymers

*Cardiac muscle and the actin-myosin system*

Regulation of contractile activity in cardiac muscle is dependent upon a cooperative interaction between thick and thin filament sarcomeric proteins that include myosin-heavy and light chains, actin, cardiac troponin-T (Tn-T) complex and tropomyosin (TM ) (see figure 1) within a fundamental unit of a sarcomere [11]. During muscle contraction, ATP hydrolysis metabolic energy is utilized as actin and myosin filaments slide past each other and produce force. The TM-Tn complex acts to inhibit the actin-myosin interaction in the resting state. With an increase of $Ca^{2+}$ in the myofilament space and binding of $Ca^{2+}$ to Tn, this inhibition is released. The hypothesis behind the steric model of activation is that $Ca^{2+}$ - Troponin Complex (TnC) induced movement of TM reverses the off state by releasing actin sites for reaction with myosin. The binding of the myosin head to actin, i.e. formation of a cross-bridge, results in a power stroke and mutual displacement of the actin and myosin filaments. Simultaneous action of many sarcomeres leads to a muscle contraction.

It was demonstrated that this interaction can change the dielectric permittivity of the sarcomeres [10], which causes a change in the SPR minimum angle. SPR measurements on sarcomeres immobilized on a gold film can provide a nearly immediate indication of attachment status of cross-bridges within intact sarcomeres in response to changing conditions. Unlike other SPR work [6], in the present study the SPR angle changes induced by contractions of sarcomers occur without binding of additional mass to the immobilized protein.

*Dipole moment and dielectric properties of tubulin dimers*

Proteins with appropriate electrical and structural characteristics (such as self-assembly into ordered networks) have been suggested as suitable building block candidates for the next generation of biological electronic devices due to their small size, minute energy requirements for conformational changes and prompt response of their state to the activating processes [12]. The cytoskeleton is the internal scaffolding that gives eukariotic cells their three-dimensional shape, effects mitosis and mobility of certain cell types and also plays a major role in intracellular transport of proteins and organelles. Tubulin is a polar protein found mainly in the cytoskeleton and especially enriched in brain and neural tissue. Based on the tubulin atomic structure and amino acid sequence, recent molecular dynamics simulations by our group and others [13] suggest this molecule possesses a permanent electric dipole moment of the order of 1700 Debye, which may change its direction as tubulin undergoes a conformational change. Under normal physiological conditions, tubulin is found as a heterodimer, consisting of two nearly identical monomers called the α- and β- tubulin each of molecular weight of about 55kDa [14]. This 110kDa dimer has the capacity to self-assemble into microtubules (MTs) which are hollow (25nm-outer diameter, 14nm inner diameter) tubes of up to micrometer lengths, and are one of the main components of the cytoskeleton. Dipole-dipole interactions among tubulin molecules are thought to be major components of the force responsible for this polymerization [15]. The cytoskeleton has also been hypothesized to be central in cellular information processing [16] and suggested as a possible candidate for implementation of biological and/or quantum computation schemes [17]. The β- tubulin monomer can bind guanosine 5' triphosphate (GTP) in which case it exists in an energy-rich form that favors polymerization (self assembly), or it can bind guanosine 5' diphosphate (GDP-tubulin) thus being in an energy-poor form (GDP-tubulin) that favors dissociation. The GDP-GTP exchange (hydrolysis) releases approximately 0.4eV per molecule and is accompanied by a conformational change [18]. This change has been modeled as resulting in a $27^o$ angle [19] between the original line connecting the centres of the α- and β- monomers and the new centre-to-centre line. These two conformational states and their associated dipole moments have been proposed as the basis for a binary system for information storage and manipulation [17,20]. Measuring the refractive index and consequently the high-frequency dielectric constant of tubulin in solution are important for the experimental determination of the dipole moment of tubulin and eventual implementation of schemes where laser light couples to stable tubulin networks effecting conformational changes and thus encoding information.

## 3. Surface Plasmon Resonance Sensor

*Experimental setup*

Figure 1 shows the experimental arrangement of our SPR sensor used for our study of sarcomeres. A helium-neon laser provided the incident illumination at 633 nm. A p-polarized light beam convergent in an angular interval was produced with an arrangement of lenses and a polarizer. A prism provided the coupling of the laser beam to the SPs that were excited in the gold film, and a multiple-channel flow cell allowed solution access to the gold film. The angular distribution of the reflected light was measured with a photodiode array and its electrical output was read with a data acquisition (DAQ) board and transferred to a PC computer. The readout rate of the DAQ board set the time resolution to 67 milliseconds. The spatial resolution was determined by the dimensions of the laser beam spot at the surface: 0.5 mm×0.3 mm. The angular resolution for the configuration used was ~1 resonant unit (RU); this angular unit, commonly used in SPR measurements, corresponds to $10^{-4}$ degrees. The change of the SPR angle by this quantity occurred when the change of the refractive index was only $10^{-6}$ [21].

For our study of tubulin molecules we used the commercial SPR-based BIAcore 3000 sensor, which furnished the additional convenience of automated injection of very small volumes of analyte solutions.

*Theoretical Model*

To obtain quantitative data, we use a five-layer model that is based on Maxwell's equations describing reflection of light from a layered system. This enabled us to calculate SPR curves, estimate SPR response from protein immobilization and estimate changes in dielectric permittivity. We consider a structure with five layers: layer 1 consists of a prism with dielectric permittivity $\varepsilon_1$=2.30, layer 2 consists of a gold film of thickness $d_2$=47 nm with complex permittivity $\varepsilon_2 = -13.2 + i1.25$, layer 3 consists of a dextran layer filled with high relaxation (HR) solution of thickness $d_3$=140 nm [22] and $\varepsilon_3$=1.78 in the case of the sarcomeres, and with buffer in the case of tubulin, layer 4 consists of our sample medium (sarcomeres or tubulin) with thickness $d_4$ and layer 5 consists of only solution. The fifth layer is assumed to be semi-infinite with respect to surface plasmon penetration depth. One can obtain the intensity reflection coefficient R for this system with the following recursive formula [23] by calculating each input impedance $Z_{in,m}$ and each layer impedance $Z_m$.

$$R = \left| \frac{Z_{in,2} - Z_1}{Z_{in,2} + Z_1} \right|^2 \qquad (1)$$

The input impedance at layer m ($Z_{in,m}$) and layer impedance ($Z_m$) are obtained from,

$$Z_{in,m} = Z_m \left[ \frac{Z_{in,m+1} - iZ_m \tan(k_{z,m} d_m)}{Z_m - iZ_{in,m+1} \tan(k_{z,m} d_m)} \right], \quad m=2,3,4 \qquad (2)$$

where,

$$Z_{in,5} = Z_5, \quad Z_m = \frac{k_{Z,m}}{\varepsilon_m k_0}, \quad k_{Z,m} = \sqrt{\varepsilon_m k_0^2 - k^2}, \quad k = k_0\sqrt{\varepsilon_1}\sin(\theta), \quad k_0 = \frac{2\pi}{\lambda} \qquad (3)$$

The wavelength of laser in vacuum is λ (633 nm and 760 nm for measurements with sarcomeres and tubulin dimers respectively). The incidence angle θ of light onto the prism-gold interface determines the component of the wave vector k that is parallel to the interface. A change in dielectric permittivity of sarcomeres or tubulin solution ($\varepsilon_4$) alters the reflection coefficient R. Once the change in the SPR angle $\Delta\theta_{SPR}$ for different media is experimentally determined, the corresponding change in the dielectric constant and refractive index ($\Delta\varepsilon_4$, $\Delta n_4$) can be calculated from equations (1-3).

The decrease of the SPR sensitivity to changes in the dielectric permittivity with depth z starting from the boundary between 3$^{rd}$ and 4$^{th}$ layers into the protein sample was calculated. This decrease can be approximated as $\sim e^{-(z/d_P)}$ and in the case of immobilized sarcomeres and λ=633nm, the characteristic penetration depth $d_p$ can be calculated to be 110 nm [23]. Approximating the diameter of a sarcomere to be ~40 nm, one can estimate that SPR measures 2-3 layers of immobilized sarcomeres. Since in this case the thickness of the protein layer exceeds the penetration depth, the latter determines the signal at saturation.

For measurements with tubulin we had λ=760nm and $d_p \geq$ 110nm. The saturation occurs when all binding sites in the dextran matrix are occupied by tubulin dimers. Approximately, this occurs, when the total mass of immobilized protein reaches that of a monomolecular layer (with thickness of ≤10nm). Thus, one can expect that saturation signal is several times lower at immobilization of tubulin compared to sarcomeres.

## 4. Methods and materials

Gold film chips (CM5) coated with carboxymethylated dextran were obtained from BIAcore (Biacore AB, Sweden). Using standard chemical activation/deactivation protocols [24], we introduced N-hydroxysuccinimide esters into the surface matrix of the chip by modifying the carboxymethyl groups with a mixture of N-hydroxysuccinimide (NHS) and N-ethyl-N'- (dimethyl-aminopropyl)-carbodiimide (EDC). These esters then form covalent bonds with the amine groups present in the ligand molecules thus immobilizing them on the surface.

The description of the preparation of mouse cardiac myofibrils are presented elsewhere [10]. For immobilization, the myofibril suspension of sarcomeric protein particles of several micrometers in diameter was brought to protein concentration of 0.7 μg/μl, pH between 4.9 and 5.1, at room temperature. Effective immobilization requires that the pH be lower than the isoelectric point of the protein; however, lowering pH below 4.9 eliminates sarcomeric function and was avoided. Sarcomeric protein concentrations between 0.5 to 0.75 μg/μl yield an optimal loading.

The temperature controller raised and maintained the temperature of the inflow at 26$^{o}$C. Functioning sarcomeres caused repeatable SPR shifts during calcium activation cycles. For sarcomeres we estimated $\varepsilon_4$=1.96 (this corresponds to protein concentration ~40% in aqueous solution [22]).

Following established protocols [25], tubulin was purified from bovine cerebra (provided courtesy of R.F. Luduena). Our SPR measurements took place at 24$^{o}$C and the time between injection and measurement was of the order of 10 s. Measurements were taken for times up to 5 minutes. Tubulin does not polymerize at 0$^{o}$C and although 10sec is adequate time for our sample of 50 μL to reach room temperature and start polymerizing, we are confident that in our measurements mainly free tubulin dimers were present and not MTs since (using spectrophotometry and monitoring the absorption curve) we had previously determined that the characteristic time for our tubulin to polymerize into MTs was of the order of 45 minutes at room temperature (data not shown) agreeing with literature [26].

5. **Results**

*Immobilization and observation of sarcomeres*

The purified myofibril preparations were immobilized on the gold film with carboxymethylated dextran surface [10]. The immobilization SPR shift ranged from 3 to 24 kRU. This value corroborates well with the estimated SPR shift value of 25 kRU at saturation using equations (1-3). As the thin filament is activated in response to an increase in [$Ca^{2+}$], the number of attached cross-bridges at dynamic equilibrium should rise, which would cause a positive shift in the SPR measurement. Flowing replacement of HR solution with pCa4.5 solution causes SPR shifts well beyond the refractive index differences between solutions themselves (figure 2). Subsequent replacement of pCa4.5 with HR returns the SPR signal back to its baseline value. The magnitude of the pCa4.5 induced SPR shift directly correlated with the amount of myofibril immobilized on the gold film. The apparent change of the dielectric permittivity $\Delta\varepsilon$ from the experimental data due to cross-bridge formation could be estimated using the 5-layer theoretical model. The observed SPR shift of 120 RU corresponds to $\Delta\varepsilon = 1.6 \times 10^{-3}$.

The SPR shift in response to increasing and decreasing [$Ca^{2+}$] was also monitored (figure 3). Flowing replacements of solutions with increasing [$Ca^{2+}$] produce a different response curve than flowing replacements of solutions with decreasing [$Ca^{2+}$]. This hysteresis effect is most noticeable between pCa6.0 and pCa5.0. Additional tests with changing concentrations [Pi], [BDM], and [ATP] affecting the relative population of attached cross-bridges at different steps of the cross-bridge cycle confirmed that the SPR response was due to cross-bridges formation [10].

*Immobilization and measurement of the dependence of tubulin refractive index on concentration*

In this setup, a 1 ng/mm$^2$ surface immobilization yields a signal of 1kRU and the laser spot size on the gold chip is 1.2 mm$^2$. As shown in figure 5, after tubulin immobilization and application of a high flow rate (20 nL/min) of running buffer to wash away any weakly-bound protein, the average response was ~4 kRU which means 4.8 ng of protein (i.e. $2.6 \times 10^{10}$ individual tubulin dimers) were captured by the dextran. A reference cell on the same chip without any immobilized tubulin was used as control and any non-immobilization relevant signals (such as due to refractive index changes) were automatically subtracted.

Electron crystallography measurements on zinc-induced tubulin protofilament sheets have shown that the tubulin heterodimer has dimensions 46 X 80 X 65 Å [27,14] so the footprint of the molecule on the surface can be between 30 nm$^2$ minimum and 52 nm$^2$ maximum depending on orientation. Using the average of these two values, it can be seen that a monomolecular layer covering the 1.2 mm$^2$ spot would require $3.0 \times 10^{10}$ individual tubulin molecules, leading us to believe that we have achieved 87% coverage, an observation corroborated by the immobilization part of the sensogram where a tendency towards saturation can be clearly seen (figure 5 D→E).

*In vitro* polymerization happens spontaneously at room temperature (also at 37°C only faster) if the protein concentration is above critical and the buffer contains adequate GTP. The critical concentration varies for different tubulin preparations. By using spectrophotometry, we determined that our tubulin started polymerizing at room temperature when concentration exceeded 1.0 mg/ml (data not shown). In order to determine the dielectric constant of tubulin we first had to be sure that the shift in SPR angle was due to the change in the refractive index of the solution floating over the gold chip and not due to further immobilization of protein or perhaps tubulin-tubulin binding (polymerization).

To address the first concern, we performed the experiment in parallel, utilizing a reference channel on the same chip but without any tubulin in it. The reference signal was automatically subtracted from the tubulin signal thus also addressing concerns related to non-specific binding to deactivated dextran. To eliminate the possibility that our signal was due to further tubulin-tubulin interactions on the surface (polymerization) we tried both below critical (0.51mg/ml) and above critical (1.7mg/ml and 5.1mg/ml) concentrations and saw return to baseline in all cases showing that in this environment tubulin was incapable of polymerization, a fact that may be due to dextran binding and/or insufficient nucleation sites. Using the sensogram of figure 5 and the theoretical model we calculated the changes of the refractive index and dielectric constant with tubulin concentration:

$$\frac{\Delta n}{\Delta c} = (1.85 \pm 0.20) \times 10^{-3} (mg/ml)^{-1} \text{ and } \frac{\Delta \varepsilon}{\Delta c} = (5.0 \pm 0.5) \times 10^{-3} (mg/ml)^{-1}$$

where $\Delta n$ and $\Delta \varepsilon$ are the changes in the refractive index $n$ and dielectric constant $\varepsilon$; $\Delta c$ is the change in concentration $c$.

As the dielectric constant and refractive index of a solution are intimately connected to the polarizability and consequently to the dipole moment of its constituents, these measurements show that SPR can be used to further elucidate the dielectric properties of 'live' proteins in solution.

## 6. Conclusions

The data presented demonstrate that SPR measurements can be used to monitor cross-bridge attachment/detachment inside sarcomeres in real time. We have also employed the SPR sensing technique to measure the refractive index and high frequency dielectric constant of a protein in solution. These methods can be extended to address the interplay between molecular binding activity and dielectric environment and can be used to decipher the nature of the relation between the two. Such work would greatly enhance our understanding of phenomena such as protein self-assembly into ordered networks like the sarcomere or the cytoskeleton, contractile function, mitosis and protein transport and would supplement biochemical approaches to shedding light on the complicated inner workings of cells.

**Acknowledgements**

We gratefully acknowledge R.F. Luduena's expert help and purification of tubulin. We would also like to thank A. Banerjee, P.M. Schwarz, M. Hook, A. Hook, J.K. Kim, and P.D. Gershon. This work is partially supported by a grant from the Telecommunications and Informatics Task Force (TITF) from Texas A&M University and a National Science Foundation grant for 'Quantum and Biological Computing' (NSF grant number 0218595). A. Mershin is partially supported by a scholarship from the A. S. Onassis public benefit foundation.

**Figure captions**

Figure 1. Schematic of the experimental setup. The prism provides the coupling of the excitation light to the surface plasmon. The polarizer P1 is used to produce the p-polarized light, since only this component interacts with the surface plasmon. The angular distribution of the reflected light intensity is detected by the photodiode array (PDA). The sample medium is injected into a small flow cell (FC) adjacent to the gold film (GF). The insert schematically presents the detail of the sensor chip with immobilized sarcomeres. In the relaxed state the tropomyosin acts to inhibit the actin-myosin interaction. With an increase of $Ca^{2+}$ in the myofilament space and its binding to troponin complex, this inhibition is released. The following binding of the myosin head to actin, i.e. formation of a cross-bridge, results in a power stroke and mutual displacement of the actin and myosin filaments (contraction).

Figure 2. A sensogram showing the immobilization of sarcomeres. Initial rise in the SPR angle corresponds to the injection of the activation solution. The following gradual increase of the SPR angular shift reflects the binding of sarcomeres onto the gold film. HR wash means injection of high relaxation solution [10].

Figure 3. The sarcomeres' reaction to changes in [$Ca^{2+}$] causes SPR shifts: the injection of pCa4.5 solution triggers a SPR shift of greater magnitude with immobilized sarcomeres than for the solutions alone [10].

Figure 4. For a fixed myofibril immobilization, this panel shows SPR shifts normalized to maximum shift versus pCa for (○)-increasing and (■)-decreasing [$Ca^{2+}$] [10].

Figure 5. BIAcore3000 sensogram for tubulin. Tubulin was immobilized on Channel 1. Channel 2 was treated identically to Channel 1 but had no tubulin. Ch.1 - Ch.2 shows tubulin signal - background. A→B: running buffer. B→C: EDC/NHS dextran-activating complex injected, note identical response of both channels. C→D: running buffer. D→E: Ch.1 shows tubulin

immobilization with a clear tendency towards saturation, Ch.2 remains at running buffer baseline level. E→F: high flow rate running buffer. Difference between Ch1. and Ch.2 shows amount of immobilized tubulin ~4000RU. F→G: ethanolamine blocking (dextran-deactivation). G→H: running buffer. H→I: 0.51mg/ml tubulin in Ch.1. Since both channels exhibit same signal, all signal is due to refractive index change, not tubulin-tubulin binding (there is slight non-specific binding to Ch.2). I→J: running buffer. J→K: 1.70mg/ml tubulin in Ch.1, all signal is due to refractive index change. K→L: running buffer, L→M: 5.1mg/ml tubulin in Ch.1, all signal is due to refractive index change. Slight noise in the forms of bumps in Ch1-Ch.2 is due to 0.5 sec delay between measurement of Ch.1 and Ch. 2 and subsequent subtraction. Bump at around 4000sec is due to the temporary presence of a bubble in the 5.1mg/ml tubulin sample.

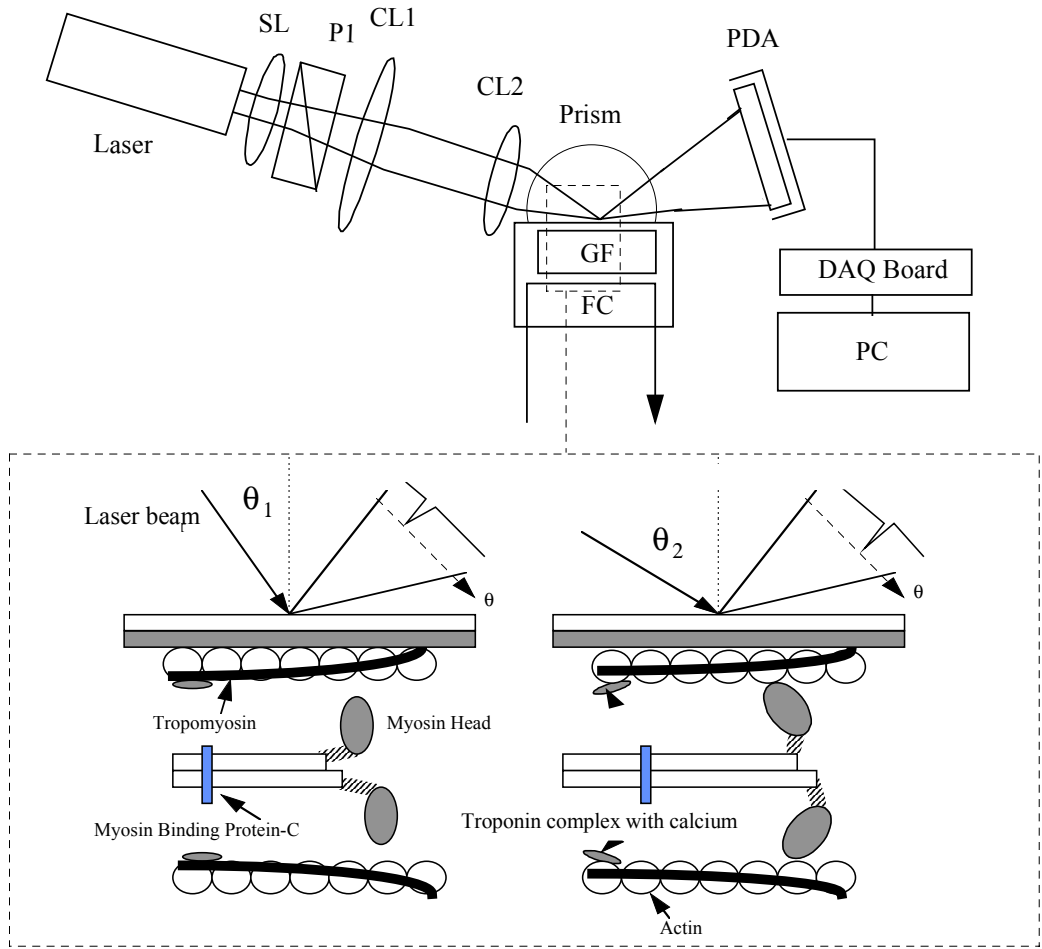

Figure 1

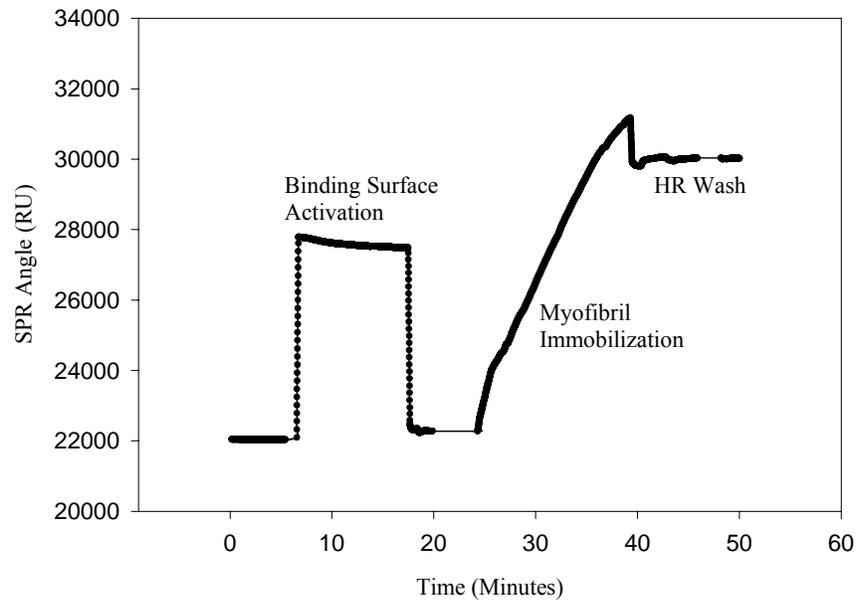

Figure 2

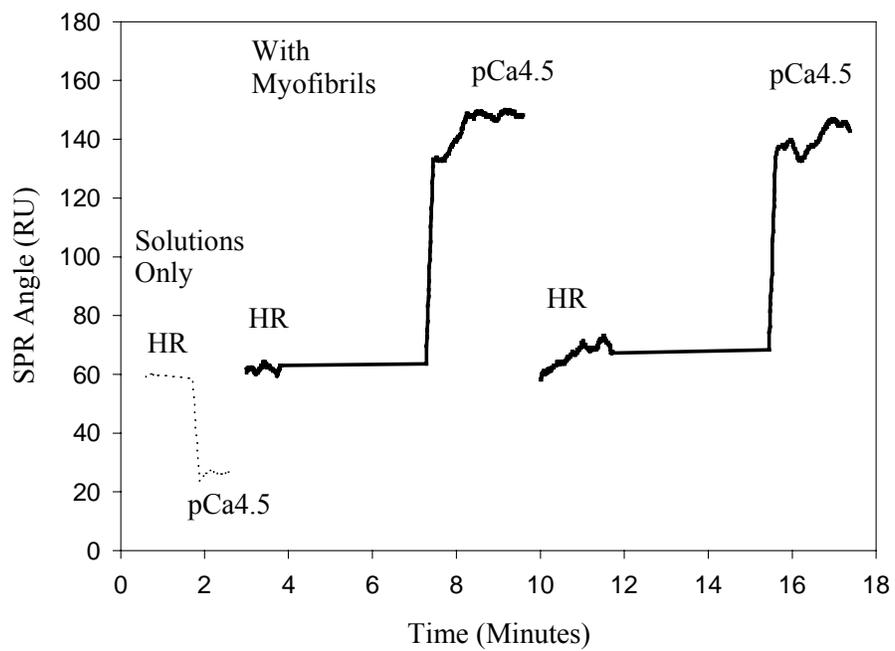

Figure 3

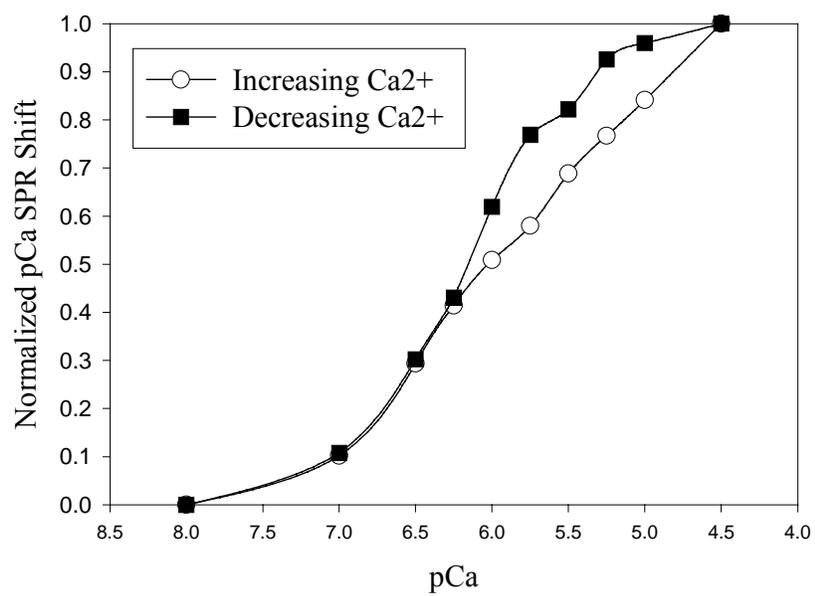

Figure 4

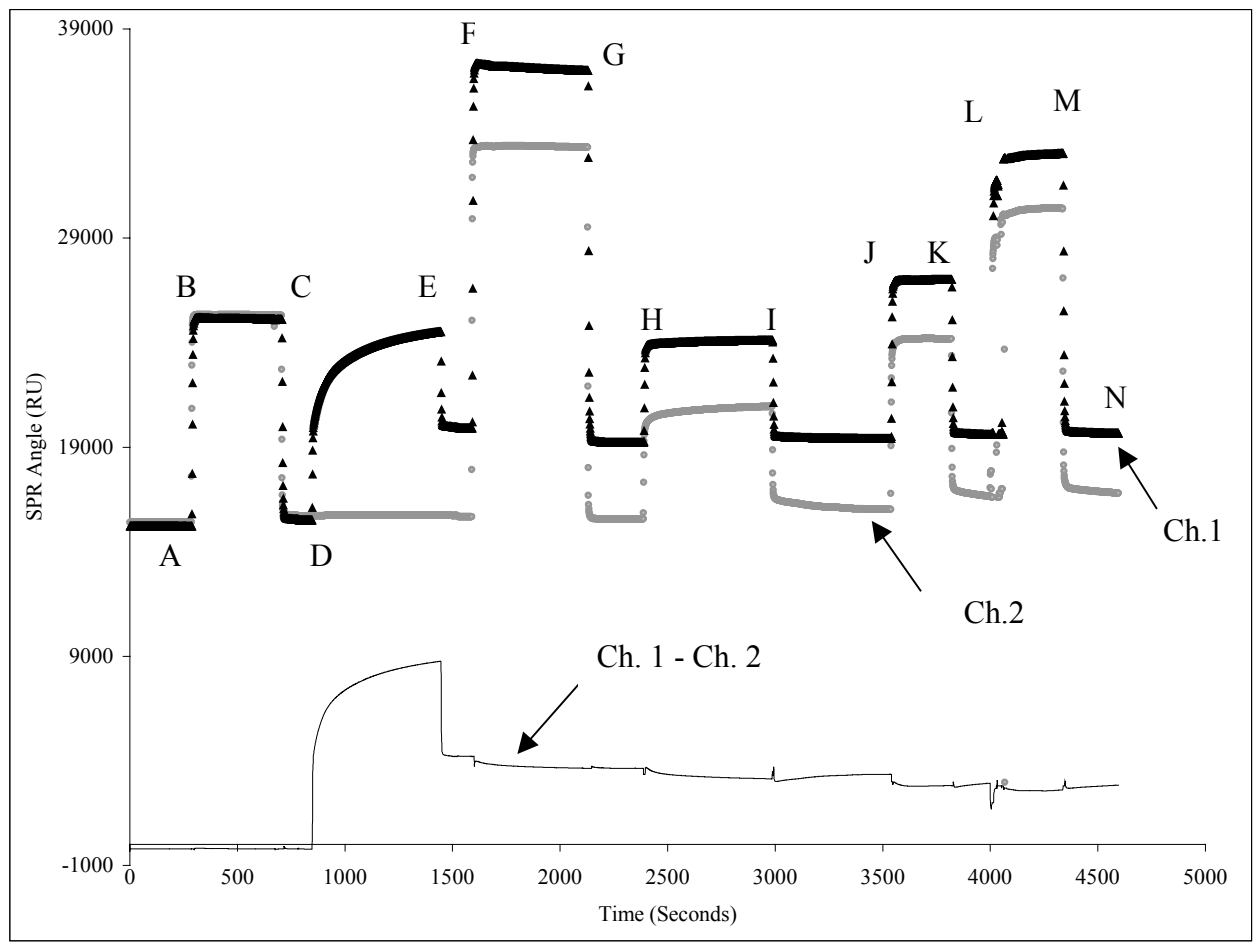

Figure 5